\documentclass{PoS}
\usepackage{graphicx}
\newcommand{\be}{\begin{eqnarray}}
\newcommand{\ee}{\end{eqnarray}}

\title{Some thoughts on critical and tricritical points and deconfinement}

\ShortTitle{Critical/deconfinement}

\author{\speaker{Edward Shuryak}\thanks{Partially supported by the US-DOE grant DE-FG02-88ER40388}\\
        Department of Physics and Astronomy\\
        Stony Brook University, Stony Brook NY 11794 USA\\
        E-mail: \email{shuryak@tonic.physics.sunysb.edu}}

\abstract{We start with a discussion of a phase diagram and
three special collision energies: (i) the first touching of the mixed phase;
(ii) the softest point; (iii) the critical point. We then proceed
to more details about the role a massless critical mode
should play in the particle interaction and the elliptic flow.
We then enter a discussion of whether high density QGP is or is
not strongly coupled, and if so what one can say about
color superconductivity domain using universality and 
data on trapped ultracold fermionic atoms.  At the end, we
briefly discuss 
 deconfinement, or more specifically on whether there are
precursors of the QCD strings at $T>T_c$, as well as possible role
of magnetically charged excitations (monopoles and dyons). 
        }

\FullConference{The 3rd edition of the International Workshop --- The Critical Point and Onset of Deconfinement --- \\
		 July 3-7 2006\\
		 Galileo Galilei Institute, Florence, Italy}

\begin{document}
\section{Mapping this talk on the phase diagram}
\begin{figure}[h]
\begin{minipage}[c]{6.cm}
 \centering 
\includegraphics[width=5.5cm]{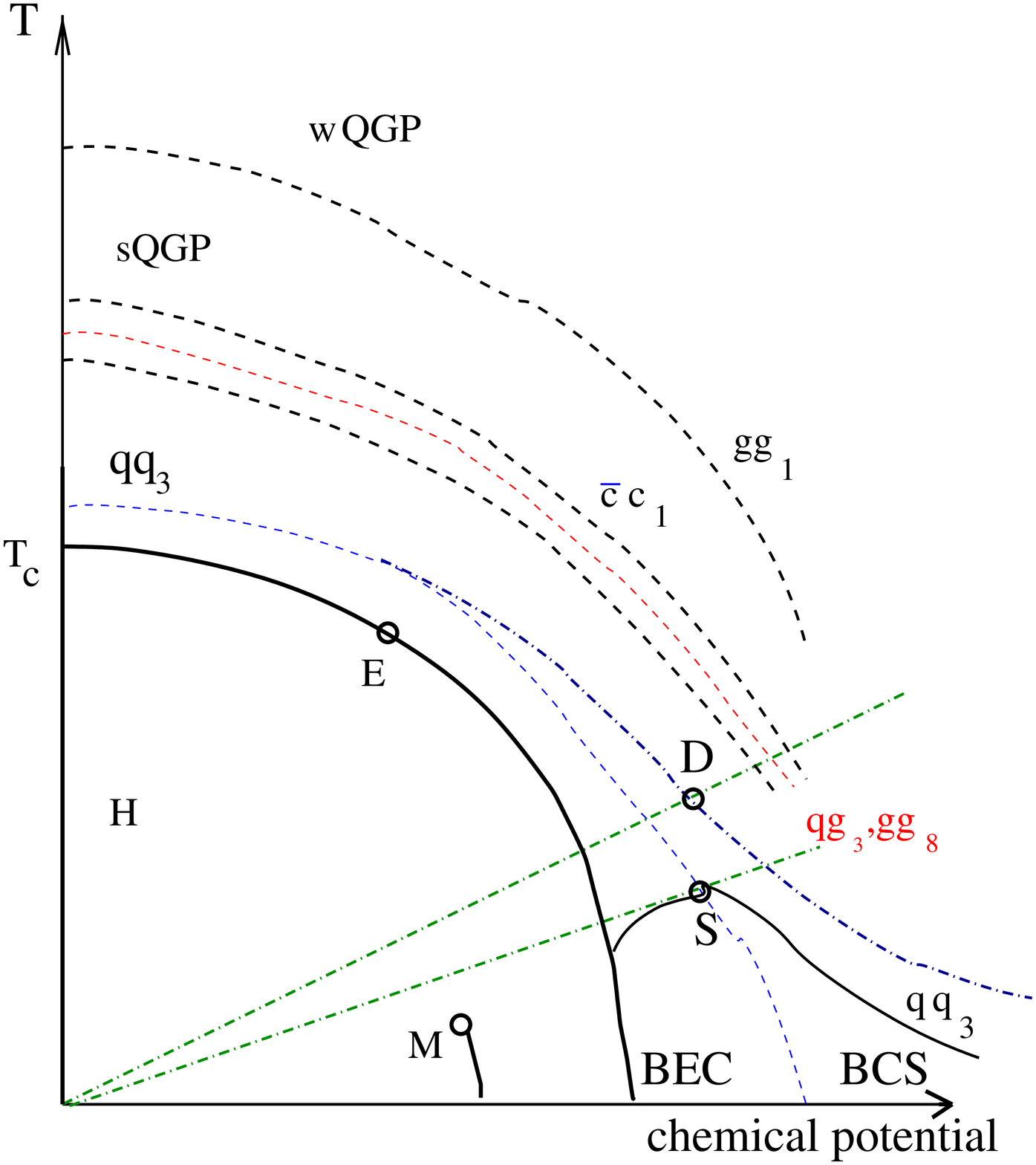}
 \end{minipage}
\hfill
\begin{minipage}[c]{6.cm}
 \centering 
\vskip .4cm
\includegraphics[width=5.5cm, angle=-90]{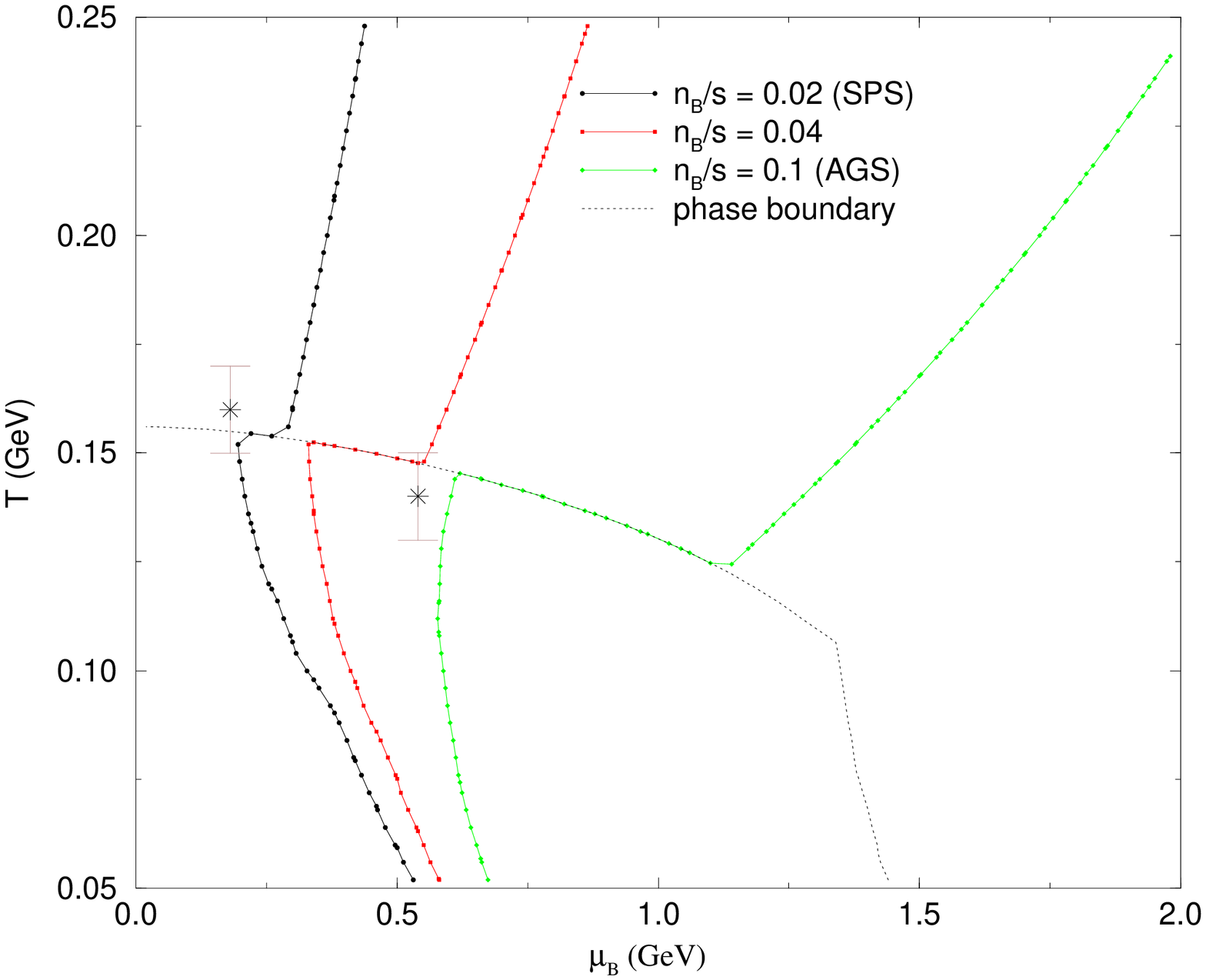}
\end{minipage}
\caption{(a)  Schematic phase diagram for QCD, in the plane baryon
  chemical potential - temperature.
M (multi-fragmentation) point is the endpoint of nuclear gas-liquid transition.
E is a similar endpoint separating the first order transition
to the right from a crossover to the left of it.
(Black) solid lines show
phase boundaries, dashed  lines are curves
of marginal stability of indicated states. 
Two dash-dotted straight lines are related with bounds from atomic
experiments we discuss in the text, they intersect with
unbinding of diquark Cooper pairs (D) and most strongly coupled point (S),
which is at the maximum of the transition line ans is
also a divider between BCS-like and BEC-like color superconductor.
(b) Example of the adiabatic cooling paths, from \protect\cite{HS}.
}
\label{fig_phase_diag}
\end{figure}

Let us
start with a presentation of our main map
of the territory, the QCD phase diagram. Its most used
form is a plane of temperature $T$ -baryonic\footnote{Of course one can
introduce different chemical potential for each quark flavor. However,
we do not need to do so in this talk.
} chemical potential $\mu$. A schematic version of it 
we will need for our purposes is shown in
Fig.\ref{fig_phase_diag}. The low-T-low-$\mu$ part is a confining
hadronic phase denoted by
H. The very high-T region is a weakly coupled quark-gluon plasma
denoted by wQGP, and the region not too high above is now known
as strongly coupled one, sQGP. The dashed lines 
introduced by Zahed and myself \cite{SZ_rethinking} show expected 
``zero binding lines'' of corresponding binary channels, which are 
also known as ``curves of marginal stability'' (CMS). (The
subscript is color representation, so $\bar c c_1$ means color singlet
charmonium, and $qq_3$ a color triplet diquarks. ) The right lower
part of the diagram is the domain of color superconductivity, 
 believed to be separated from QGP by a second order
transition line. The place where two transition line meet
is called the ``triple'' point:
here all three basic phases of QCD -- confined,
QGP and Color Superconductivity  -- can coexist.
The issue where is it at the phase diagram depends on  
  the {\em maximal pairing} possible: 
indeed one is often asked whether 
it can 
 be reached via heavy ion collisions. We will address recent progress
on this in section \ref{sec_CS}.

The point $M$ is the endpoint of
liquid-gas transition of nuclear matter, and
we believe there is another  critical endpoint
denoted by $E$, which separates 
the cross over (to the left of it) from the first order transition
to the right of it. The exact location of point E is subject
to intense work on the lattice, which is yet to converge to as
definite value. At $E$ there is a second order transition,
which must be accompanied by specific phenomena associated
with infinite correlation length: we will return to it in section
\ref{sec_critical} where we will argue   that this should
dramatically increase attractive mean field potentials for baryons
and non-Goldstone mesons. 
In particular, we
 expect a downward shift of the mass of 
vector mesons
$\rho,\omega$ to be greatly increased in its vicinity,
which is directly accessible via dilepton experiments.
For Goldstone modes -- the pions --
 we predict that their interaction is instead becoming stronger
and more repulsive. There are observable
 implications of these effects for
collective  flows  of pions and nucleons.

 As  other talks at this conference,  
 we focus our discussion 
 toward rather poorly explored region in the middle of the map,
where forthcoming high-density (HD) RHIC runs and eventually
FAIR experiments would take place. It is widely
expected
that those may locate
the critical endpoint $E$. Moreover, in section \ref{scan} we will show
as the collision energy grows from around few GeV/N there are
two more
 special energy points. The first corresponds to energy when
the ``initial point'' reached the phase transition line for the
first time. The second
(known as the ``softest point'') is when it starts venturing into the
QGP domain.

Of course, thermodynamical quantities are not the only ones which
one would like to know.
Many other practically important quantities can be calculated
perturbatively
in QGP,
such as dilepton and photon production
rates. 
{\em Transport properties} like  viscosity or particle diffusion
constant are especially   important, because they define
the micro length defining applicability of the
approach. (Unfortunately,
it is next to impossible calculate those on the lattice.)

It is in this area where recent revolution happened: both RHIC data
as well as some theory development have lead to a conclusion
 \cite{discovery_workshop}
that QGP is in fact in a ``strongly coupled'' regime, named sQGP,
and is a near-perfect liquid rather than 
a weakly interacting gas. Fascinating 
connections between theory of sQGP and other strongly coupled systems
 -- string theory, ultracold atoms, e/m strongly coupled plasmas --
 have lead to a tremendous progress in the last few years.
We will not however address ``strong coupling'' issues in this talk
(except in relation to triple point).

Another long-standing problem is the famous problem of
confinement.
More specifically, we would like to understand what  happens
 at the deconfinement region, at $|T-T_c| \ll T_c$. In section
 \ref{monopoles} we will discuss some new ideas about its precursors
from the plasma side. 
\section{Three special energy points for heavy ion collisions }
\label{scan}

 The phase diagram is a
theoretical concept, appropriate for the infinite matter in
equilibrium. The idea of ``traveling'' from one point to another
includes the assumption that all transitions happen slowly,
 at a time scale 
long enough to get new equilibrium established. Clearly, this is
not so during initial period of the collisions, in which large
amount of entropy is produced by very non-equilibrium phenomena:
therefore (unlike for textbook heat engines)
 one can map only the cooling part of the cycle.

Those are
described by ``adiabatic paths'': an early example\footnote{
It comes from a decade-old  paper Hung and myself \cite{HS}
devoted to hydrodynamics of radial flow at SPS.} 
of such paths is shown in 
Fig.\ref{fig_phase_diag}(b), for a particular
resonance gas plus bag-model QGP thermodynamics
with the first order transition.
 Since the baryon number is
also
conserved, the ratio of baryon to entropy densities $n_B/s$ 
remains constant along the line: those values are shown in the figure.
I show it to emphasize their zigzag shapes: the reason is
fixed   $n_B/s$ in H and QGP phases do not match. 
Continuity demands that
(in spite of continuous
expansion!) there is a ``reheating'' part of the zigzag
hiding behind the phase boundary. 
(Two data points correspond to 
 experimentally extracted
 ``chemical freezeout points'' for different
collision energies: more recent points
of the kind will be  shown in Fig.\ref{fig_phase_diag_CS} below.)

\begin{figure}[t]
\centering
\includegraphics[width=7cm]{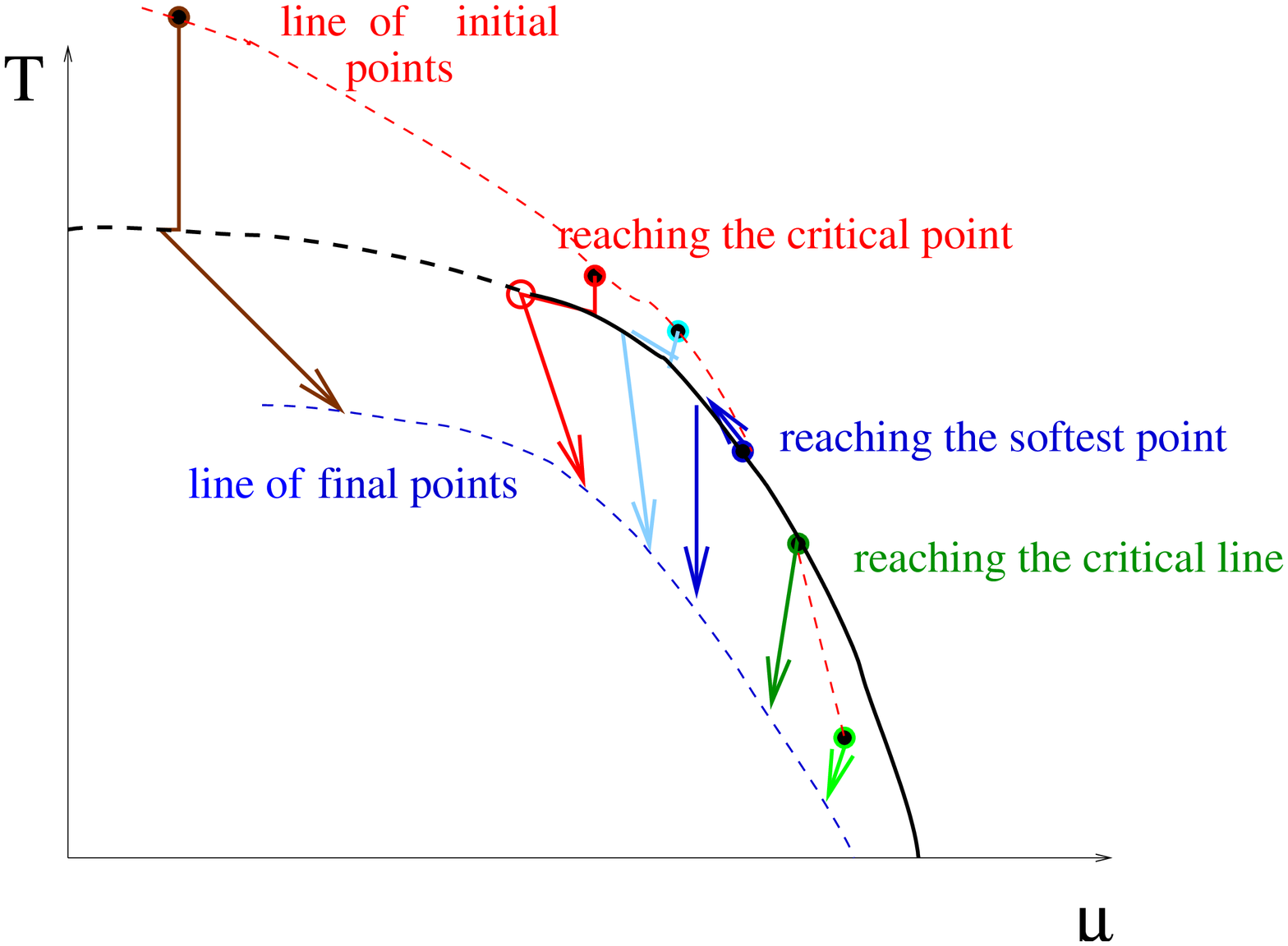} 
\includegraphics[width=5cm]{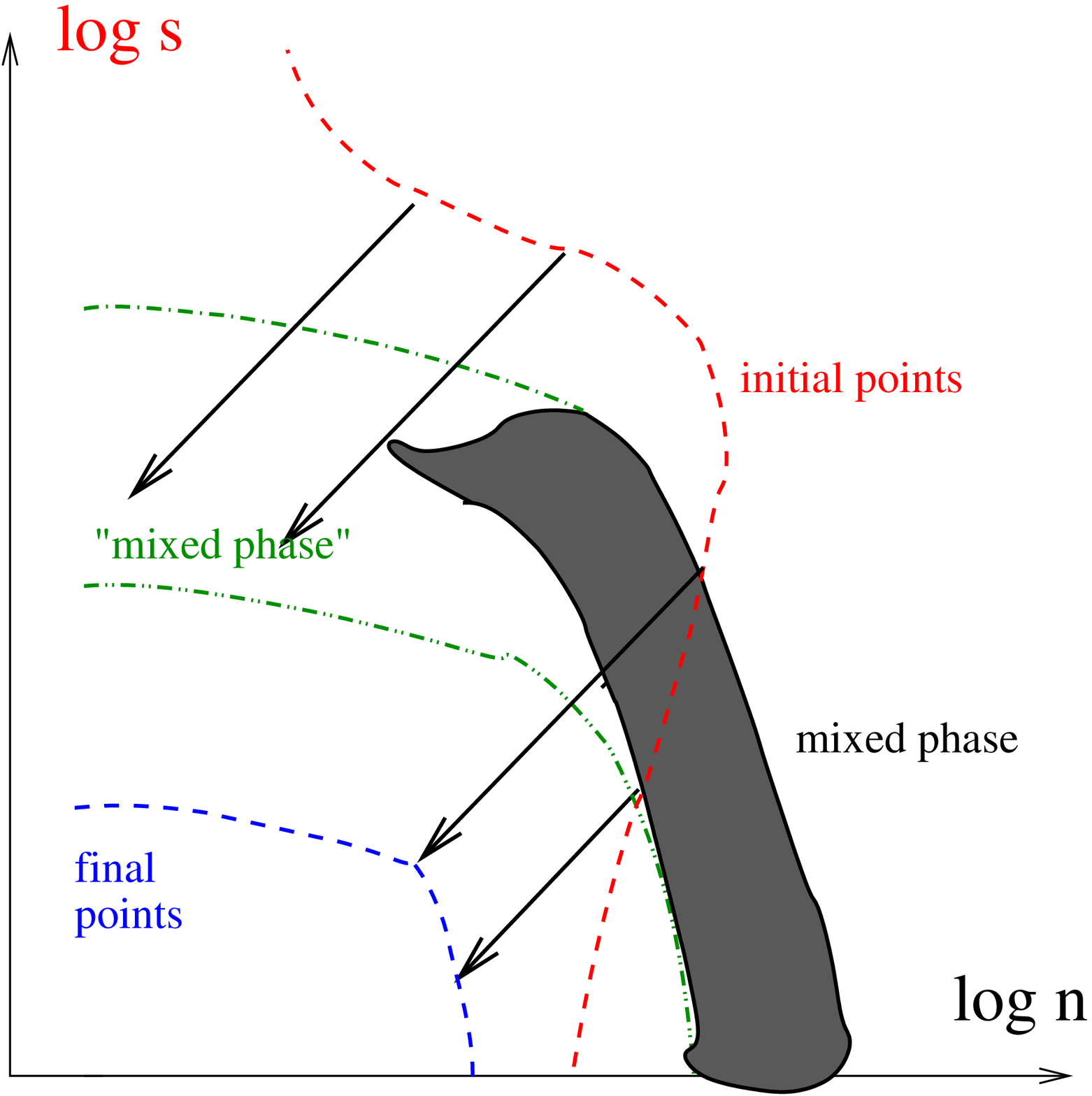}
\vskip 0.4cm
\caption{\label{fig_phase_diag_events}
Schematic view of the cooling paths on (a) $T-\mu$ and (b)
$log(s)-log(n)$ diagrams. 
In (a) those paths are
 zigzag-shaped lines with
arrows, extended from the (red dashed) line of the initial
points, to (blue dashed) line of the kinetic freezout.
In (b) the same lines are a set of parallel straight lines.
}
\end{figure}

Let us now look at  Fig.\ref{fig_phase_diag_events}
which shows  more sophisticated phase diagram,
indicating special collision points specifically
relevant for (high-density) HD-RHIC--FAIR energy domain. 
The upper (red) dashed line 
corresponds to the  ``initial'' points reached in the
collision\footnote{ 
Although its definition is rather vague  -- equilibrium is ``nearly
reached'' because ``most of the entropy''
is already generated --  changing it 
would presumably move this line a bit, but not the cooling paths
stemming from it (the zizgzags with arrows) themselves.}.
 Those cooling paths,
 distinguished via  $s/n_B$ value, go through the transition line,
if it is on its way, toward the ``kinetic freezout line'' shown by
the lower (blue) dashed line. 

At very low energy we are at the right lower
end of the diagram: and both the initial and final lines are
in hadronic (nuclear) matter phase.
As the energy of the collision grows, one reaches 
{\it the first singular
point}, at which
 {\bf the initial point touches the phase boundary}. 
The {\it second} one is when {\bf the initial point separates from
the ``mixed phase'' into the QGP phase}. This point was 
called {\bf the softest point} in \cite{HS}, because here $p/\epsilon$
ratio
is believed to reach its minimum. It corresponds to longer-lived
fireball expansion.

As lattice results show, moving to the left along the phase boundary
one eventually find
a ``rapid crossover'' regime at small $\mu$,
 with the second-order
critical endpoint $E$.  The first-order-related 
zigzags are supposed to change near  the critical point
into a different ``focusing'' behavior, suggested in \cite{SRS} and
detailed in \cite{Nonaka}.
 
Another way to look at the phase diagram,
perhaps more appropriate for our applications, is to transfer
all quantities from  ($T,\mu$) representation 
to their thermodynamical conjugates, the entropy and baryon densities
($s,n$). 
A schematic picture of such plot is shown in  
Fig.\ref{fig_phase_diag_events}(b). In this case the cooling curves are
extremely simple: as we use log-log scale, they are just a set
of parallel lines with the unit slope
\be log(s)-log(n)=const\ee
But ``the law of conservation of difficulties''
cannot be ignored: now the phase boundaries become very complicated
instead. Furthermore, in such variables
the ``mixed phase'' domain
(hiding inside the 1-st order line before)
 now opens up and gets visible (the dashed region). If the
whole transition would be first order, the whole region between
(green) dash-dotted lines would be a mixed phase. The shadows
gray area corresponds to ``true'' mixed phase.
As in the previous figure, we also indicated the set of initial
points  by a (red) dashed line, and a ``final'' line by (blue) dashed
line blow. All three special points indicated above are
seen on this plot nicer.

 Returning to the experimental program, for HD-RHIC and FAIR,
one may say that their main objectives should be {\bf location
of all three collision energies} corresponding to those three points.
As also emphasized by other talks, we dont know where
the critical point is, but most likely it is around
$E_{Lab}=20-40\, GeV*A$. The ``softest point'' is perhaps
at $10-20\, GeV*A$: maybe it is related with 
the (so far  unexplained ) ``horn'' in  $K/\pi$ ratio 
observed by the NA49 collaboration 
at the lowest SPS energies due to different expansion
time. The ``first touch'' point
I think is around  $E_{Lab}\sim 5-6\, GeV*A$; the
corresponding experimental hits are perhaps the  sudden change in
energy dependence of
radial and elliptic flows.

\section{The critical mode and its effect on flow}
\label{sec_critical}
Let me first remind that thermodynamics (and also of
the masses of the related modes) follows from
the Landau-Ginzburg function, which was
discussed in detail in the second paper \cite{SRS}.
Without quark masses (in the chiral limit) one has a
so called tricritical point, which becomes simply critical
when finite $u,d$ quark masses are included. 
As is well known, Goldstones -- the pions -- are massive 
anywhere, including the critical point\footnote{Where $m_\pi^2\sim m^{4/5}$
which is only slightly differs from $m^1$ in Gellman-Oaks-Renner
relation in the vacuum. Below we will ignore this small correction 
and treat pion mass as unchanged from the vacuum value. }.

The real difference should come from the {\em ``critical mode''}
which (in terms of its ``screening mass'') must gets massless at the
critical point. And, at this point, massless means
really massless, for the real world with nonzero
quark masses. Although we will refer to it  as ``sigma''
field below, actually the critical mode is a mixture between
scalar and vector $\sigma-\omega$ mesons\footnote{At nonzero baryon
  density, C-parity is broken and such mixing is allowed.}

  Stephanov, Rajagopal and myself \cite{SRS} were the first to
 propose experimental search for the QCD (t)critical point in
  heavy ion
collisions, 
by varying the energy of and looking for ``non-monotonic'' signals.
The first specific signal  proposed in that paper was an increased 
{\em event-by-event fluctuations} near it, reminiscent of critical
opalescence well known near other 2-nd order phase transitions. 
 The second proposal of \cite{SRS} was the so called ``focusing
effects'' already mentioned above. Both are quite subtle signals.
In this talk I would like to   discuss my later suggestions
\cite{Shu_masslesssigma} of few
(hopefully) more robust signals of the critical point.

\begin{figure}[t]
\centering
\includegraphics[width=8cm]{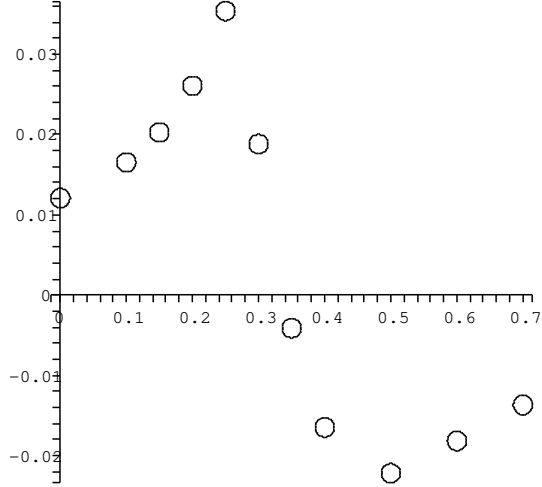}
\vskip 0.4cm
\caption{\label{fig_crit_pion_pot}
Effective potential for a pion at rest $Re[V_{eff}(p=0)]$ [GeV]
induced by sigma resonance, as a function of the sigma mass $M_\sigma$
[GeV].
}
\end{figure}

All non-Goldstone particles (such as baryons and e.g. vector mesons)
have effective mean field potentials in hadronic phase, a significant
part of which is the attractive sigma exchange. If one 
take any particular
 model of those (e.g. Walecka model for nuclear forces) and put
naively sigma mass to zero, huge effect follows, which cannot possibly 
be realistic. It happens because the
nuclear potential  is in fact a {\em highly tuned small difference}
 of
two large terms. This also makes any predictions difficult.
Yet there is little doubt that infinite-range attraction
will increase the potentials.

One observable consequence of that can be extra large mass shifts for
$\rho,\omega$ mesons
\be \label{eq_Rmassshift}
  \Delta V_{\rho,\omega}={2\over 3}\Delta V_N=-{2\over 3}n_s \large({{g_s^*}^2 \over
  {m_\sigma^*}^2}-{g_s^2 \over m_\sigma^2} \large) \ee
These shifts were experimentally seen in nuclear matter, where
$\omega$ mass shifts by about 14\% \cite{metag}, as well as  in
heavy ion 
collisions e.g. in peripheral AuAu collisions at RHIC,
where  STAR
collaboration \cite{STAR_Fachini} finds about 10\% shift
of $\rho$.  
Extra mass shift near the critical point should be of the same
order of magnitude.

For pions the effect should be different. First of all, they 
must interact
with sigma with derivatives, to protect their Goldstone masses,
thus there is no significant effect of large distance (small momentum)
sigmas on them. On the contrary: if sigma mass (now a real one, not
the screening) shifts so that $m_\sigma<2m_\pi$, sigma gets unable 
to decay to pions and also strong repulsion in the $\pi\pi$
channel will follow. Thus very light sigma implies a $repulsive$
effect on pions (evaluated in \cite{Shu_masslesssigma}, see Fig.\ref{fig_crit_pion_pot}
).

 Now, if
the nucleon-nucleon and nucleon-non-Goldstone-meson interaction
 gets much more attractive near the critical
point than it is elsewhere, 
this should reduce collective flows of nucleons, both
their radial and elliptic components. At the same time, as the
pion-pion interaction gets more repulsive, one should expect 
the opposite effect  in the pion flows. 

Sudden reduction
of the baryonic flow at low SPS energy was reported by NA49:
whether it is a real effect we hope to learn soon from HD-RHIC
runs.

\section{The location of a triple point and (strongly coupled) color superconductivity}
\label{sec_CS}
{\em Color superconductivity} (CS) in dense quark matter
in general appears due to attractive interaction in certain scalar
diquark channels. More specifically, three mechanisms
of such attraction were discussed in literature
are: (i) the electric Coulomb interaction 
(see e.g.\cite{Bailin:1984bm}) 
(ii) instanton-induced 't Hooft interaction 
\cite{RSSV,ARW} and the 
magnetic interaction \cite{Son:1999uk}. The
instanton mechanism have re-activated the field, because it
 have demonstrated for the first time
 that large (and thermodynamically significant)
pairing may actually appear in cold quark matter. Since
forces induced by small-size instantons are 
not affected by screening that much, they can be much stronger
than (i), leading to
 gaps  about 2 orders of magnitude larger
 $\Delta\sim 100\, MeV$ \cite{RSSV,ARW}.

 However such gaps are rather uncertain because
 they are
so large\footnote{ 
Additional argument \cite{RSSV} why one should believe such large gaps:
 the $same$ interaction but in $\bar
q q$ channel is responsible for chiral symmetry breaking, producing the gap
(the constituent quark mass) as large as 350-400 MeV.
Furthermore, in the {\it two-color} QCD, the so called Pauli-Gursey
symmetry  relates these two condensates directly.} that they
may fall outside of applicability range
of  the usual BCS-like mean field theory. Unfortunately,
an analytic theory of  strong pairing is still
in its infancy.
 The main idea of this letter
is  to get around this 
``technical'' problem by using the universality arguments
and  an appropriate atomic data.

 The interaction (scattering) 
of two quarks
is maximally  enhanced (to its unitarity limit) if there is a marginal 
state in the diquark channel,   a bound state
with near-zero binding or  a virtual state at small positive
energy. In atomic systems such situations are generically called
a ``Feshbach resonance'', tuned by external magnetic field.
 Transition between these two possibilities is 
reflected in specific behavior of the condensation, known as
BEC-to-BCS transition. In the middle, with resonance at
exactly zero binding, the interaction is
at its maximum, limited by
unitarity (for the relevant s-wave).

In quark-gluon plasma  the existence of marginally bound states
of quarks and gluons 
and their possible role in liquid-like behavior at RHIC has been
pointed out in \cite{SZ_rethinking}. 
For any hadronic states 
one can argue that they dissolves at large $T$ or $\mu$,
and thus existence of some lines of zero binding
on the phase diagram are unavoidable.
The issue was so far studied only for small
$\mu$ and high $T$ relevant for RHIC heavy ion program:
and indeed there are theoretical, lattice and experimental
evidences that e.g. charmonium ground state does not melt till
about $T\sim 400\, MeV$. 

 In \cite{Shuryak_CS} I had explored consequences of the idea that
there is a $diquark$ marginal binding line, closely
trailing the phase boundary line on the phase diagram.
If this is the case (which  I cannot  prove at this time),
it must cross the CS critical line, separating the CS region
into a BCS-like and BEC-like, see right lower part of 
 Fig.\ref{fig_phase_diag}(a).
 Lines of zero binding of a $qq$ pair  
 starts at $\mu=0$ at the temperature $T_{qq}$ very close to
the critical line $|T_{qq}-T_c|\ll T_c$: the reason for that
is that effective color attraction in $qq$ state is only $1/2$
of that in mesons such as charmonium. 
 The lines associated with (color triplet) diquark
$qq_3$ are
the
lower (blue) dashed and dash-dotted lines: below the former  
diquark binding is below zero and above the later it does not exist at
all,
even as a Cooper pair.

\begin{figure}[h]
\includegraphics[width=8cm]{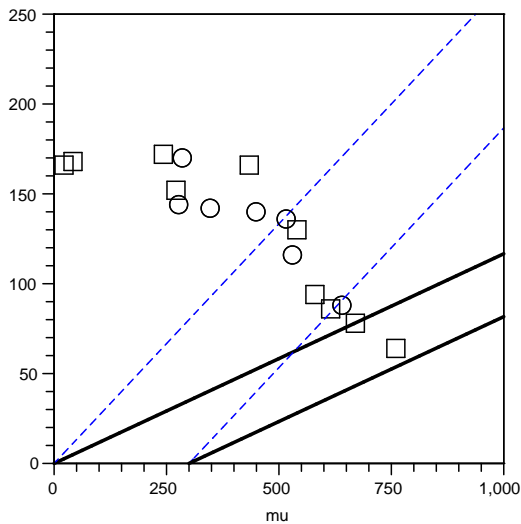}
\caption{Datapoints are for compilation
 of fits to chemical freezeout parameters
from different experiments according to
\protect\cite{chemical}.
The squares and circles are for fits at mid-rapidity and all
particles, respectively. Two solid lines are the phase transition lines
with the quark effective mass $M_1=0$ and $100\, MeV$, two dashed
  lines show pair unbinding lines for
 the same masses.
}
\label{fig_phase_diag_CS}
\end{figure}

How can one relate  atomic and  quark systems, if at all?
The way it can be done is due to the so called ``universality''
of the system. As the scattering length gets large $a\rightarrow
\infty$,
it cannot enter the answers any more. As a result, 
there remain so few parameters on which the answer may
depend\footnote{
The next parameter of scattering amplitude, the effective range,
is about 3 orders of magnitude smaller than interparticle distance
for trapped atoms, and thus completely irrelevant.
Although similar parameter for
quarks are not that small, we will assume it does not
affect the universal results too much.
}
on that those  can be absorbed by selection of the appropriate units.

The 50-50 mixture of two ``spin'' states is characterized
by the density $n$ and the particle mass $m$. 
Quantum mechanics adds $\hbar$ to the list of possible quantities,
and so one has only 
3 input parameters, which can be readily absorbed by
selecting proper units of length, time and mass. Thus the pressure
(or mean energy) at
infinite and zero $a$ can
only
be  related by some universal
 numerical  constant 
 $p_\infty / p_0= (1+\beta)$. We do not know how to get 
its value from any theory
(other than quantum Monte-Carlo simulations or other brute force
methods), 
 but it has been  measured 
experimentally (e.g. by the very size of the trapped system). 
The same should hold for transport properties: e.g.
viscosity of such universal gas
 can only be  $\eta=\hbar n \alpha_\eta$ where $\alpha_\eta$
is some universal coefficient \cite{GSZ}. 
Similarly, the critical temperature 
must be simply proportional to the Fermi energy 
\be \label{eqn_alpha_def}
T_c=\alpha_{T_c} E_F\ee
with the universal constant $\alpha_{T_c}$.

Experimental progress in the field of strongly coupled
 trapped fermionic atoms is quite spectacular; unfortunately
 this author is certainly unqualified to  go into its discussion.
 Let me just mention one paper, which killed 
remaining doubts about superfluidity: 
 a discovery by the MIT group of
the quantized vortices, neatly
organized
into the usual lattice
\cite{MIT_vort}.
 We will however need only  the
  information about the $value$ of the transition temperature. 
Duke group lead by J.E.Thomas
(Kinast et al) have for some time studied collective vibrational
modes of the trapped system. Their frequencies in strong coupling
regime are well predicted by hydrodynamics and universal
equation of state, with little variation with temperature.
Their damping however show
 significant $T$-dependence: in fact Kinast et al
\cite{kinast_damping} have found two distinct transitions
in its behavior. 
The lowest break in damping
is interpreted as the phase transition to superfluidity, it
corresponds to \be \alpha_{T_c}={T_c\over E_F}=.35\ee
 where $E_F$ stands for
the Fermi energy of the ideal Fermi gas at the center of the
trap.
 In another (earlier) set of experiments
 \cite{kinast_cv} there has also been found a change in the specific
heat,   at   $\alpha_{T_c}=.27$. In spite of some
numerical difference between these two values,
 the Duke group indicates
that both are related to the same phenomenon\footnote{ They
  do not provide any
error bars on the value of $T$, as the absolute value of the temperature
is obtained in rather indirect calibration procedure. The reader
may take the spread of about 20\% as an estimate of uncertainties
involved.}. 
At another  temperature  
\be \alpha_2={T_2\over T_F} \approx 0.7-0.8 \ee
 the behavior of the damping
visibly changes again. Kinast et al
 interpret it as a transition to a regime where
not only there is no $condensate$ of atomic pairs, but
even the  pairs themselves are melted out.

4. 
There are of course important differences between quarks and atoms.
First, quarks have not only spin but also flavor and color, so
there are $3*N_f$ more Fermi surfaces. However, in the first approximation one
may focus only at one pair of them (say u-d quarks with red-blue
colors)
which are actually paired.
Second, atoms are non-relativistic while quarks are 
in general relativistic 
and in matter may have some complicated dispersion laws.
Since the gaps are large, one cannot use a standard argument
that close to Fermi surface only Fermi velocity is important.
Nevertheless, we will
 assume that quark quasiparticles
have dispersion laws which can be approximated by a simple
quadratic form
\be E(p)=M_1+{p^2\over 2 M_2}+... \ee 
where dots are for $O(p^4)$ terms we ignore.  In matter
$M_1$ and $M_2$ need not be the same. 
 It then implies that e.g. the relation between the critical $T$ and
chemical potential
should read
\be  T_c= \alpha_{T_c}(\mu/3-M_1)\ee
and similarly for the second point.
(The factor 3 appears because baryon number is counted per baryon, not quark.)

The $M_2$ can be used
to set the units as explained above, 
and we will not actually need its value.
 The $M_1$ is  needed, but since it makes   a simple shift of
the chemical potential, it can be eliminated by differentiation.
Thus we get a prediction of the slope of the critical
line 
\be dT_c/d\mu= \alpha_{T_c}/3\approx 0.1\ee  
The intersection of CMS for diquarks
with the SC critical line (the point S (strong) in our phase diagram   
Fig.\ref{fig_phase_diag}(a)) should
thus be at this line, at which the boundary of superconducting
phase
crosses with the zero energy of the bound state.
The second critical point associated with disappearance of pairs
(identified with the (blue) dash-dotted line and point D
in Fig.\ref{fig_phase_diag}(a)) should thus be at the line with the
slope
$\alpha_2$. 

In order to plot the line on the phase diagram one needs the value
of the $M_1$, which unfortunately is not known. To set the
 {\em upper bound}
one may simply take $M_1=0$ and draw
two straight lines pointing to the origin,
see Fig.1(b). 
As $M_1$ grows, the lines slide to the right,
as is shown by another line with a
 (randomly chosen) value $M_1= 100 \, MeV$.

Finally, we turn to ``realistic'' phase diagram, with
numerical values extracted from experiment.
 We know for sure  that matter is released at
the so called chemical freezeout lines indicated by points in
Fig.\ref{fig_phase_diag}(b). These points are
the ends of adiabatic cooling paths.  Unfortunately we do not
know how high above them those curves start at a given collision
energies. However it is general expected that this line more or less
traces the critical line, being few MeV below it, into
 the hadronic phase. One may then conclude that the upper limit
on $T_c$ of CS is about 70 MeV
(intersection of the upper solid lines with the freezeout line). The
disappearance of pairing (dashed lines) is thus expected below
$T_2=150\, MeV$. 

Finally, comparing these results with theoretical expectations
and experimental capabilities, we conclude that  
(i) if there is a strongly coupled CS,
its critical temperature should definitely be below $T_c<70\, MeV$.
This maximal value is amusingly close  to
what was obtained from 
the instanton-based calculations \cite{RSSV,ARW};
(ii) therefore {\bf no heavy ion collisions
can reach the CS domain}, 
even at the maximal coupling. However, a penetration into the region
$T= 100-150\, MeV, \mu=500-600\, MeV$ in which
non-condensed bound diquarks may exist, I think is quite likely,
both in the HD-RHIC runs  and in future GSI facility FAIR;
(iii) if that happens, one may think of some further
 uses of universality,
e.g. about relating the transport properties in both systems.
One may in particularly ask whether the universal
 viscosity extracted from vibrations
of trapped atoms (like that
in \cite{GSZ}, but at appropriate $T$)
 can or cannot describe the hydrodynamics
of the corresponding heavy ion collisions.

\section{sQGP and deconfinement}
\label{monopoles}

The \cal{N}=2 SUSY YM (``Seiberg-Witten'' theory) 
is a working example of confinement due to 
condensed monopoles\cite{SW}.
If it is also true for QCD,  at $T\rightarrow T_c$
magnetic monopoles must become light
 and weakly interacting at large distances due to U(1) beta function.
Then the Dirac condition forces
 electric coupling $g$  be large (in IR).

Recent lattice data have revealed a puzzling behavior
of static $\bar Q Q$ potentials, which I call
``postconfinement''. At $T=0$ we all know that a potential between
heavy quarks is a sum of the Coulomb and a confining $\sigma(T=0) r$ 
potential. At deconfinement $T=T_c$ the Wilson or Polyakov lines with
a static quark pair
has  vanishing string tension; but this is the free energy
$exp(-F(T,r))=<W>$. Quite shockingly, if one calculates the $energy$
or $entropy$ separately (by $F=E-TS$, $S=-\partial F/\partial T$)
one finds 
 a force between $\bar Q Q$ to be more than twice
  $\sigma(T=0)$ till rather large distances. The total energy 
added to a pair is surprisingly large, it
reaches about $E(T=T_c,r\rightarrow\infty)=3-4\, GeV$, and the
entropy as large as $S(T=T_c,r\rightarrow\infty)\sim 10$. Since this energy of ``associated matter''
is about 20 times larger than $T$, any separation of two static quarks
must be extremely suppressed by the Boltzmann factor exp(-E/T).
(As $T$ grows, this phenomenon disappears,
and thus it is obviously related to the phase transition itself.)

Where all this energy and entropy may come from in the
 deconfined phase? 
The most likely reason is a long (thus energetic)
 QCD string of a compicated shape (thus a lot of entropy)
connecting two static quarks. Such strings can be explained by 
a ``polymerization'' of gluonic quasiparticles in sQGP \cite{LS}.

Let us now add a twist to this story related with magnetic
excitations,
the monopoles\footnote{Recall that they appear naturally if
there is an explicit Higgs VEV breaking of the color group.
We cannot discuss in detail a QCD setting: the reader may
simply imagine a generic finite-$T$ configuration with
a nonzero mean $<A_0>$, an  adjoint Higgsing leaving
$N_c-1$ U(1) massless gauge fields. These U(1)'s corresponds to
magnetic charges of the monopoles. In AdS/CFT language one may simply
considered $N_c$ branes to be placed not at exactly the same point
in the orthogonal space.}.
 According to t'Hooft-Mandelstamm scenario, confinement is supposed to be due to monopole
condensation. Seiberg-Witten solution for the \cal{N}=2 SYM is an
example of how it is all supposed to work: it has taught us that
as one approaches the deconfinement transition
the electrically charged particles -- quarks and gluons --
are getting heavier while monopoles gets lighter and more numerous.
Although I cannot go into details here, we do have hints from
lattice studies of monopoles and related observables that
this is happening in QCD as well.
 
Let us now think what
all of it means for the sQGP close to $T_c$. Even at classical level, it means that
one has a plasma with both type of charges -- {\it electric and magnetic} --
at the same time, with the former dominant at large $T$ and
the latter dominant close to $T_c$.  

A binary dyon-dyon systems have been studied before , but not
manybody ones.
The first numerical studies of such systems
(by molecular dynamics) are now performed by
(Stony Brook student) Liao and myself \cite{LS_monopoles}. We found
that a monopole can be trapped by an electric static dipole, both
classically and quantum mechanically. 

We also found that classical
plasma containing magnetic monopoles is sufficient  to generate
 electric 
flux tubes\footnote{Those are dual to magnetic flux tubes in
solar classical plasmas.}, or the QCD strings. The reason is
 monopoles scatter from the electric
flux tube back into plasma, compressing it.
Whether monopoles  are condensed or not is not that crucial.
Such solutions are in fact dual to known magnetic flux tubes
in solar plasma.

Are there bound states of electric and magnetic quasiparticles?
Yes, there are a lot of them. A surprise is that
even  finite-$T$ instantons
can be viewed as being made of $N_c$ selfdual dyons~\cite{Kraan:1998kp},  attracted to
each other  pairwise, electrically and magnetically.
  Not only such baryons-made-of-dyons have the same moduli space
as instantons,  the solutions can be obtained
vis very interesting AdS/CFT brane
construction \cite{Lee:1997vp}. Many more exotic bound states
of those are surely waiting  to
be discovered.


\begin{thebibliography}{99}

\bibitem{SZ_rethinking}E.~V.~Shuryak and I.~Zahed,
  Phys.\ Rev.\ C {\bf 70}, 021901 (2004)
  [arXiv:hep-ph/0307267].
Phys.\ Rev.\ D {\bf 70}, 054507 (2004)
  [arXiv:hep-ph/0403127].
\bibitem{discovery_workshop}
  E.~V.~Shuryak,Prog.Part.Nucl.Phys.53:273-303,2004, hep-ph/0312227
  Nucl.\ Phys.\ A {\bf 750}, 64 (2005).
M.~Gyulassy and L.~McLerran,
  Nucl.\ Phys.\ A {\bf 750}, 30 (2005)
  [ nucl-th/0405013].
\bibitem{HS}
  C.~M.~Hung and E.~V.~Shuryak,
   ``Equation of state, radial flow and freeze-out in high energy heavy ion
  Phys.\ Rev.\ C {\bf 57}, 1891 (1998)
  [arXiv:hep-ph/9709264].

\bibitem{SRS}
M.~Stephanov, K.~Rajagopal and E.~Shuryak,
Phys.\ Rev.\ Lett.\  {\bf 81}, 4816 (1998)
[hep-ph/9806219].
``Event-by-event fluctuations in heavy ion collisions and the {QCD}  critical point,''
Phys.\ Rev.\  {\bf D60}, 114028 (1999)
[hep-ph/9903292].

\bibitem{Nonaka}
  C.~Nonaka and M.~Asakawa,
  ``Hydrodynamical evolution near the QCD critical end point,''
  arXiv:nucl-th/0410078.
\bibitem{Shu_masslesssigma}
 E.~Shuryak,
  ``Interactions between hadrons are strongly modified near the QCD
  (tri)critical point,''
  arXiv:hep-ph/0504048.
\bibitem{metag} CBELSA/TAPS Collaboration (D. Trnka et al.). Apr 2005. 4pp. Phys.Rev.Lett.94:192303,2005; nucl-ex/0504010
\bibitem{STAR_Fachini} 
 P.~Fachini,
  J.\ Phys.\ G {\bf 30}, S735 (2004)
  [arXiv:nucl-ex/0403026].
\bibitem{Bailin:1984bm}
D.~Bailin and A.~Love,
``Superfluidity And Superconductivity In Relativistic Fermion Systems,''
Phys.\ Rept.\  {\bf 107}, 325 (1984).
\bibitem{RSSV}
R.~Rapp, T.~Sch{\"a}fer, E.~V.~Shuryak and M.~Velkovsky,
``Diquark Bose condensates in high density matter and instantons,''
Phys.\ Rev.\ Lett.\  {\bf 81}, 53 (1998)
[hep-ph/9711396].
\bibitem{ARW}
M.~Alford, K.~Rajagopal and F.~Wilczek,
``QCD at finite baryon density: Nucleon droplets and color
superconductivity,''
Phys.\ Lett.\  {\bf B422}, 247 (1998),
[hep-ph/9711395].

\bibitem{Son:1999uk}
D.~T.~Son,
`Superconductivity by long-range color magnetic interaction in  high-density quark matter,''
Phys.\ Rev.\  {\bf D59}, 094019 (1999)
[hep-ph/9812287].

\bibitem{Shuryak_CS}E.~V.~Shuryak,
  ``Locating strongly coupled color superconductivity using universality and
  experiments with trapped ultracold atoms,''
  arXiv:nucl-th/0606046.

\bibitem{MIT_vort}M. W. Zwierlein, J. R. Abo-Shaeer, A. Schirotzek, C. H. Schunck and W. Ketterle,Nature 435, 1047-1051 (23 June 2005) 


\bibitem{GSZ}  B.~A.~Gelman, E.~V.~Shuryak and I.~Zahed,
  ``Cold Strongly Coupled Atoms Make a Near-perfect Liquid,''
Phys. Rev. A 72, 043601 (2005), nucl-th/0410067.

\bibitem{kinast_cv} J. Kinast et al, Science 25, jan.2005
\bibitem{kinast_damping}J. Kinast et al, cond-mat/0502507)
\bibitem{chemical}
A. Andronic, P. Braun-Munzinger, J. Stachel, 
Nucl.Phys.A772:167-199,2006,
 nucl-th/0511071

\bibitem{SW}
N.~Seiberg and E.~Witten,
  ``Electric - magnetic duality, monopole condensation, and confinement in N=2
  supersymmetric Yang-Mills theory,''
  Nucl.\ Phys.\ B {\bf 426}, 19 (1994)
  [Erratum-ibid.\ B {\bf 430}, 485 (1994)]
  [arXiv:hep-th/9407087].

\bibitem{LS} J.~Liao and E.~V.~Shuryak,
Nucl.Phys.A, in press, arXiv:hep-ph/0508035.
  Phys.\ Rev.\ D {\bf 73}, 014509 (2006)
  [arXiv:hep-ph/0510110].

\bibitem{LS_monopoles} J.~Liao and E.~V.~Shuryak, in progress

\bibitem{Kraan:1998kp}
  T.~C.~Kraan and P.~van Baal,
  ``Exact T-duality between calorons and Taub - NUT spaces,''
  Phys.\ Lett.\ B {\bf 428}, 268 (1998)
  [arXiv:hep-th/9802049].

\bibitem{Lee:1997vp}
  K.~M.~Lee and P.~Yi,
  ``Monopoles and instantons on partially compactified D-branes,''
  Phys.\ Rev.\ D {\bf 56}, 3711 (1997)
  [arXiv:hep-th/9702107].
\end{thebibliography}
\end{document}